\documentclass[useAMS,usenatbib]{mn2e}
\usepackage{multicol}
\usepackage{multirow}
\usepackage{natbib}

\usepackage{graphicx}
\usepackage{txfonts}
\usepackage{psfig}
\usepackage{subfigure}
\usepackage{natbib}
\usepackage{german}
\title[Comparing simulations and observations in RCW 120]
{Comparing simulations of ionisation triggered star formation and observations in RCW 120}
\author[Walch et al.]
{S. Walch$^{1}$\thanks{E-mail: walch@ph1.uni-koeln.de}, 
A.P.~Whitworth$^{2}$, T.G.~Bisbas$^{3}$, D.A.~Hubber$^{4,5}$, R.~W\"unsch$^{6}$\\
$^{1}$University of Cologne, Z\"ulpicher Str. 77, 50937 Cologne, Germany \\
$^{2}$School of Physics \& Astronomy, Cardiff University, 5 The Parade, Cardiff CF24 3AA, Wales, UK \\
$^{3}$Department of Physics \& Astronomy, University College London, Gower Place, London WC1E 6BT, UK\\
$^{4}$Technical University Munich, Excellence Cluster Universe, Boltzmannstr. 2, 85748 Garching, Germany\\
$^{5}$University Observatory Munich, Department of Physics, Ludwig-Maximilians-University Munich, Scheinerstr.1, 81679 Munich, \\
$^{6}$Astronomical Institute, Academy of Sciences of the Czech Republic, Bocni II 1401, 141 31 Prague, Czech Republic\\
Germany. }

\begin{document}

\date{Accepted . Received 2015, March 12; in original form }

\pagerange{\pageref{firstpage}--\pageref{lastpage}} \pubyear{2015}

\maketitle

\label{firstpage}

\begin{abstract}
Massive clumps within the swept-up shells of bubbles, like that surrounding the galactic H{\sc ii} region RCW 120, have been interpreted in terms of the Collect and Collapse (C\&C) mechanism for triggered star formation. The cold, dusty clumps surrounding RCW 120 are arranged in an almost spherical shell and harbour many young stellar objects. By performing high-resolution, three-dimensional SPH simulations of H{\sc ii} regions expanding into fractal molecular clouds, we investigate whether the formation of massive clumps in dense, swept-up shells necessarily requires the C\&C mechanism. In a second step, we use RADMC-3D to compute the synthetic dust continuum emission from our simulations, in order to compare them with observations of RCW 120 made with APEX-LABOCA at 870 $\mu\rm{m}$. We show that a distribution of clumps similar to the one seen in RCW 120 can readily be explained by a non-uniform initial molecular cloud structure. Hence, a shell-like configuration of massive clumps does not imply that the C\&C mechanism is at work. Rather, we find a hybrid form of triggering, which combines elements of C\&C and Radiatively Driven Implosion (RDI). In addition, we investigate the reliability of deriving clump masses from their 870 micron emission. We find that for clumps with more than 100 ${\rm M}_\odot$ the observational estimates are accurate to within a factor of two and that, even at these long wavelengths, it is important to account for the radiative heating from triggered, embedded protostars.
\end{abstract}

\begin{keywords}
Galaxies: ISM - ISM: nebulae - H{\sc ii} regions - bubbles - dust - Hydrodynamics - Stars: formation 
\end{keywords}

\section{Introduction}

The possibility of star formation triggered by ionizing feedback from young, massive stars has been explored for several decades. From a theoretical point of view, two main triggering mechanisms have been suggested: Collect and Collapse (C\&C), and Radiation Driven Implosion (RDI). 

The C\&C mechanism was first analyzed by \citet{Elmegreen1977}. In this mechanism, an expanding H{\sc ii} region sweeps up a layer of cold gas and dust beyond the ionization front \citep[e.g.][]{Dale2007}, and this shell eventually becomes gravitationally unstable due to the growth of perturbations along its surface \citep{Elmegreen1994, Whitworth1994b, Dale2009, Wunsch2010}. One argument in favour of the C\&C mechanism is that it is predicted to spawn massive fragments \citep{Whitworth1994a}, and hence it affords the possibility of forming massive stars sequentially. The study by \citet{Thompson2012} of massive, young stellar objects (YSOs) associated with bubbles suggests that triggering could be responsible for the formation of between 14\% and 30\% of all massive stars.

In contrast, RDI \citep{Sandford1982, Kessel2003, Bisbas2011, Haworth2012} triggers star formation by compressing pre-existing cold, but otherwise gravitationally stable, molecular cloud cores. \citet{Elmegreen1995} were the first to show that bright rims are caused by ionization erosion around {\it pre-existing} turbulent clumps. Observational and theoretical studies \citep[e.g.][]{Gritschneder2009, Gritschneder2010} suggest that RDI leads to star formation in the tips of pillar-like structures, for example as seen in the Eagle nebula \citep{White1999}. \citet{Tremblin2012} have conducted a similar study but focus on the formation of cometary globules, which form if the turbulent pressure is high. \citet{Bisbas2011} have published a detailed numerical study of RDI in initially stable Bonnor-Ebert spheres. 

The nearby H{\sc ii} region RCW 120 is one of the best studied H{\sc ii} regions in the Galactic plane. As observed with Spitzer at $8\,\mu\rm{m}$ \citep{Churchwell2006}, it appears to be an almost perfectly round bubble with a well defined ionization front. \citet{Zavagno2007} and \citet{Deharveng2009} have combined Spitzer and 2MASS data with observations at 870 ${\rm \mu}$m and 1.2 mm, to analyze the star formation associated with RCW 120. They conclude that star formation in the shell of RCW 120 has probably been triggered by a combination of triggering mechanisms. \citet{Zavagno2010} study the YSO properties with {\it Herschel} PACS and SPIRE. They confirm the existence of a YSO with mass $M_{_\star}\!=8\;{\rm to}\;10\,{\rm M}_\odot$ in one of the condensations, and identify a number of lower mass (0.8 to $4\,{\rm M}_\odot$) Class 0 sources within the shell. Massive clumps have also been found within the swept-up shells of other bubbles, e.g. Sh 104 \citep{Deharveng2003} and RCW 79 \citep{Zavagno2006}, and it has been suggested that the C\&C mechanism, which invokes the formation of massive clumps via shell fragmentation, could be at work. For all of these bubbles, the mass of the swept-up shell is in agreement with the expected swept-up mass \citep{Anderson2012}, as estimated using a simple model of an H{\sc ii} region expanding into a uniform-density ambient medium \citep[e.g.][]{Spitzer1978,Whitworth1994a}.


\citet{Walch2012, Walch2013} show that clumpy, shell-like structures like that seen in RCW 120 are probably attributable to pre-existing density structures in the natal molecular cloud. During the expansion of the H{\sc ii} region and the collection of the dense shell, the pre-existing density structures are enhanced and lead to a clumpy distribution within the shell. The masses and locations of the swept-up clumps depend on the fractal density structure of the molecular cloud, through the parameters $n$ and $\rho_0$ \citep[see Section \ref{sec2} and][]{Walch2012, Walch2013}. Subsequently, the clumps grow in mass, and at the same time they are overrun and compressed by the H{\sc ii} region, until they become gravitationally unstable and collapse to form new stars. Due to the formation of massive clumps, it is possible that there is a second generation of massive star formation. This is a hybrid triggering scenario, which combines elements of both the C\&C and RDI mechanisms. In this paper we show that the distribution of massive clumps formed around an H{\sc ii} region, which expands into a structured molecular cloud, is in good agreement with observations of massive clumps around e.g. the Galactic H{\sc ii} region RCW 120.

The plan of the paper is the following. In section \ref{sec2} we describe the algorithm used to generate initial fractal molecular clouds, and the numerical method used to evolve them, including the treatment of ionizing radiation. In section \ref{sec3} we describe the resulting H{\sc ii} regions and the modeled synthetic 870 $\mu$m observations. We discuss the shell and clump masses inferred from the synthetic observations in section \ref{sec4}, and compare them with the true masses. Our main conclusions are summarized in section \ref{Conclusions}. 

\section{Initial conditions and numerical method}\label{sec2}

\subsection{The generation of a fractal molecular cloud}\label{fractal}
The initial three-dimensional fractal density structure is constructed using an algorithm based on Fourier Transformation. The algorithm has three main input parameters, (i) the 3D power spectral index $n$, where $P(k)\propto k^{-n}$, (ii) the random seed ${\cal R}$ used to generate a particular cloud realisation, and (iii) the density scaling constant $\rho_0$ (see below). We populate the integer modes $k=1$ to 128 along each Cartesian axis $(x, y, z)$, where $k=1$ corresponds to the linear size of the cubic computational domain in one dimension. According to \citet{Stutzki1998}, the box-coverage dimension, $\mathcal{D}$, of a fractal structure embedded in three-dimensional space is related to the power spectrum by 
\begin{equation}
\mathcal{D}= 3- \frac{(n-2)}{2}
\end{equation}  
\citep[see also][]{Federrath2009}. Thus, defining $\mathcal{D}$ is equivalent to defining the power spectral index $n$. Here, we choose setups with $\mathcal{D}=2.4$, in agreement with observations of molecular clouds in the Milky Way \citep{Falgarone1991, Vogelaar1994, Stutzki1998, Lee2004, Sanchez2005}. This corresponds to $n = 3.2$.

After constructing the density fluctuations in Fourier space, and applying a Fast Fourier Transform to generate $\rho_{_{\rm FFT}}(x,y,z)$ on a $128^3$ grid, the resulting density field is scaled exponentially, using a scaling constant $\rho_0$:
\begin{equation}
  \rho(x,y,z)=\exp{\left(\frac{\rho_{_{\rm{FFT}}} }{\rho_0} \right)}\,.
\end{equation}
$\rho(x,y,z)$ has a log-normal density probability density function (PDF) and a clump mass distribution in agreement with observations \citep[as described in][]{Shadmehri2011}. In particular, for a given spectrum of density fluctuations, changing $\rho_0$ allows us to adjust the width of the density PDF, i.e. the variance $\sigma$ of the log-normal distribution, whilst leaving the underlying topology of the density field unchanged. In Figure \ref{Figure 1} we show the resulting density PDFs for the two setups we discuss in this paper. Both clouds have equal mass (see section 2.2), the same fractal dimension $\mathcal{D}=2.4$, and the same random seed $\mathcal{R}$, but two different values of $\rho_0$.

Before populating the computational box with SPH particles, we shift the point of maximum density to the center of the computational domain, and position the ionizing star there. Then we partition the computational box with a $128^3$ grid, compute the mass in each cell of the grid, and apportion each cell the corresponding number of SPH particles, distributed randomly within the cell. Finally we cut out a sphere with radius $R_{_{\rm MC}}$, centered on the ionizing star. 
The caveat of the fractal density setup is that the particles all have no initial velocities.
This assumption neglects possible flows towards/away from the source of ionising radiation, which might change the dynamics of the expanding HII region.

\subsection{Initial conditions}\label{SEC_IC}

\begin{figure}
\includegraphics[width=90mm]{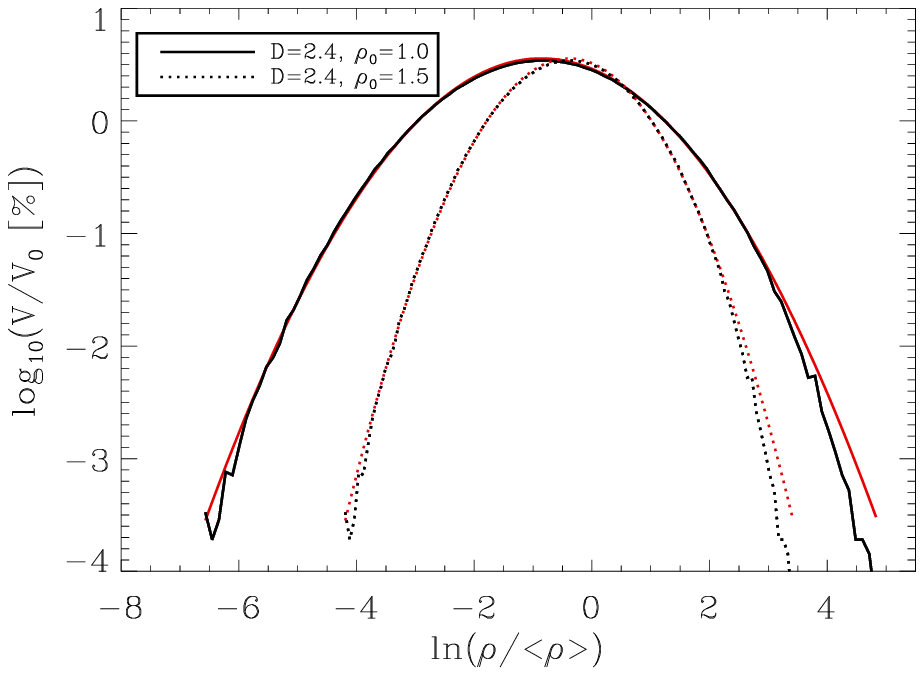} 
\caption{Volume-weighted density PDFs for fractal clouds having the same fractal dimension $\mathcal{D}=2.4$ and random seed $\mathcal{R}$, but different density scaling parameters: $\rho_0=1.5$ (dotted black line; {\it Run 1}) and $\rho_0=1.0$ (full black line; {\it Run 2}). Both distributions have been fitted with a log-normal (red lines).}
\label{Figure 1}
\end{figure}

We choose a cloud with total mass of ${\rm M}_{_{\rm MC}}=10^4\,{\rm M}_\odot$, and radius of $R_{_{\rm MC}}=5.0\,\rm{pc}$. This results in a mean density of $\bar{\rho}= 5.42 \times 10^{-21}\,\rm{g \,cm}^{-3}$, or equivalently $\bar{n}=1380\,\rm{cm}^{-3}$ for molecular gas having mean molecular weight $\mu=2.35$. The gas is initially isothermal at $T_{_{\rm MIN}}=30\,{\rm K}$. In this paper, we discuss two simulations, both of which result in a shell-like structure very similar to the one observed in RCW 120. Apart from the scaling parameter $\rho_0$, the initial clouds are identical, i.e. their density fields have the same topological structure. The simulation with $\rho_0=1.5$ (narrower density PDF in Fig. \ref{Figure 1}) is called {\it Run 1}, and the simulation with $\rho_0=1.0$ (broader PDF in Fig. \ref{Figure 1}) is called {\it Run 2}. By fitting their density PDFs using a $\chi$-squared minimisation method, we can estimate the standard deviation of the logarithmic density, viz. $\sigma_1=0.88$ ({\it Run 1}), and $\sigma_2=1.31$ ({\it Run 2}).

We can relate $\sigma$ to the conditions produced in turbulent gas by introducing the scaling relation between $\sigma$ and the turbulent Mach number derived by \citet{Padoan1997}, \citet{Padoan2002} and \citet{Federrath2008},
\begin{equation}
\sigma^2= \rm{ln}\left(1+b^2 M^2 \right)\, ,
\end{equation}
with $b \simeq 0.5$. According to this relation, the density field in {\it Run 1} corresponds to a molecular cloud with Mach 2.2 turbulence; and {\it Run 2} corresponds to Mach 4.3 turbulence. 

In both {\it Run 1} and {\it Run 2}, a star emitting hydrogen-ionising photons at rate $\dot{N}_{_{\rm LyC}}=10^{49}\,{\rm s}^{-1}$ is placed at the center of the cloud. This corresponds approximately to an O7.5 ZAMS star with a mass of $\,\sim\! 25\,{\rm M}_\odot$ and a surface temperature of $\,\sim\! 40,000\,{\rm K}$ \citep[][their Table 2.3]{Osterbrock2006}. Thus, the ionizing source is more powerful than the central star within RCW 120, which is estimated to be an O8.5 to O9 star emitting $\dot{N}_{_{\rm LyC}}=10^{48.04 \pm 0.1}\,{\rm s}^{-1}$ \citep{Zavagno2007}.

\subsection{Numerical method}

We use the SPH code \textsc{Seren} \citep{Hubber2011}, which is extensively tested and has already been applied to many problems in star formation \citep[e.g.][]{Walch2011, Bisbas2011, Stamatellos2011}. The ionizing radiation is treated with a HEALPix-based adaptive ray-splitting algorithm, which allows for optimal resolution of the ionization front in high resolution simulations \citep[see][]{Bisbas2009}. The implementation is based on the On-The-Spot approximation \citep{Zanstra1951, Osterbrock1974, Spitzer1978}, which is valid if the hydrogen number density is sufficiently high; this is usually the case, in particular in the vicinity of the ionisation front. The ionisation front is located along each ray by assuming radiative equilibrium. We employ the standard SPH algorithm \citep{Monaghan1992}. The SPH equations of motion are solved with a second-order Leapfrog integrator, in conjunction with a block time-stepping scheme. Gravitational forces are calculated using an octal spatial decomposition tree \citep{Barnes1986}, with monopole and quadrupole terms and a Gadget-style opening-angle criterion \citep{Springel2001}. We use the standard artificial viscosity prescription \citep{Monaghan1983}, moderated with a Balsara switch \citep{Balsara1995}. In both simulations we use $2.5 \times 10^6$ SPH particles. Thus, each particle has a mass $m_{_{\rm SPH}}= 4.0 \times 10^{-3}\,{\rm M}_\odot$ and the minimum resolvable mass is $\sim 0.4\,{\rm M}_\odot$ \citep{BB1997}.

The temperature of ionized gas particles well {\it inside} the ionisation front is set to $10,000\,{\rm K}$. The temperature of neutral gas particles well {\it outside} the ionisation front is given by a barotropic equation of state that mimics the gross thermal behaviour of protostellar gas \citep[][their Fig. 4]{Masunaga1998},
\begin{equation}
T(\rho)=T_{_{\rm MIN}} \left\{1+\left(\frac{\rho}{\rho_{_{\rm CRIT}}}\right)^{(\gamma-1)}\right\}\,,
\end{equation}
with $T_{_{\rm MIN}}=30\,{\rm K}$, $\rho_{_{\rm CRIT}}=10^{-13}\,{\rm g}\,{\rm cm}^{-3}$, and $\gamma=5/3$. Thus, for densities below $\rho_{_{\rm CRIT}}$, the gas is approximately isothermal at $30\,{\rm K}$, and for densities above $\rho_{_{\rm CRIT}}$, the gas is approximately adiabatic with adiabatic index $\gamma=5/3$ ($\gamma$ has the monatomic value, because, although the gas is primarily molecular hydrogen, it is too cold for the internal degrees of freedom of the hydrogen molecules, even the rotational ones, to be excited). The choice of $T_{_{\rm MIN}}=30\,{\rm K}$ is somewhat arbitrary, and was initially chosen simply because it agrees with the estimate in \citet{Deharveng2009}. However, it agrees well with the mean {\it dust}-temperature calculated a posteriori using RADMC-3D (see Section \ref{SEC:Massestimate}). The temperatures of SPH particles within one local SPH smoothing length of the ionisation front are forced to vary smoothly between the temperature of the ionised gas on one side and the temperature of the neutral gas on the other; this is required to avoid a numerical instability.

We use a new sink particle algorithm to capture forming protostars \citep{Hubber2013}. Sinks are introduced at density peaks above $\rho_{_{\rm SINK}}=10^{-11}\,\rm{g}\,\rm{cm}^{-3}$, provided that the density peak in question is at the bottom of its local gravitational potential well. Since $\rho_{_{\rm SINK}}\!\gg\!\rho_{_{\rm CRIT}}$, a peak that is converted into a sink is always well into its Kelvin-Helmholtz contraction phase. Once created, a sink has a radius of twice the SPH smoothing length at $\rho_{_{\rm SINK}}$, i.e. $R_{_{\rm SINK}}=20\,{\rm au}$. A sink then accretes mass smoothly from the surrounding SPH particles within $R_{_{\rm SINK}}$, over many dynamical timescales, but transfers their angular momentum to SPH particles just outside $R_{_{\rm SINK}}$ \citep[see][for more details]{Hubber2013}. This procedure ensures that the masses and locations of sink particles are essentially independent of the arbitrary parameters of sink creation and evolution, $\rho_{_{\rm SINK}}$ and $R_{_{\rm SINK}}$. Sinks are identified as protostars, and their luminosity can be included in the radiative transfer models produced with \textsc{Radmc-3D} in the post-processing step (Section \ref{radmc3d}).

\section{Results}\label{sec3}

\subsection{SPH simulations}
\begin{figure*}
\includegraphics[width=190mm]{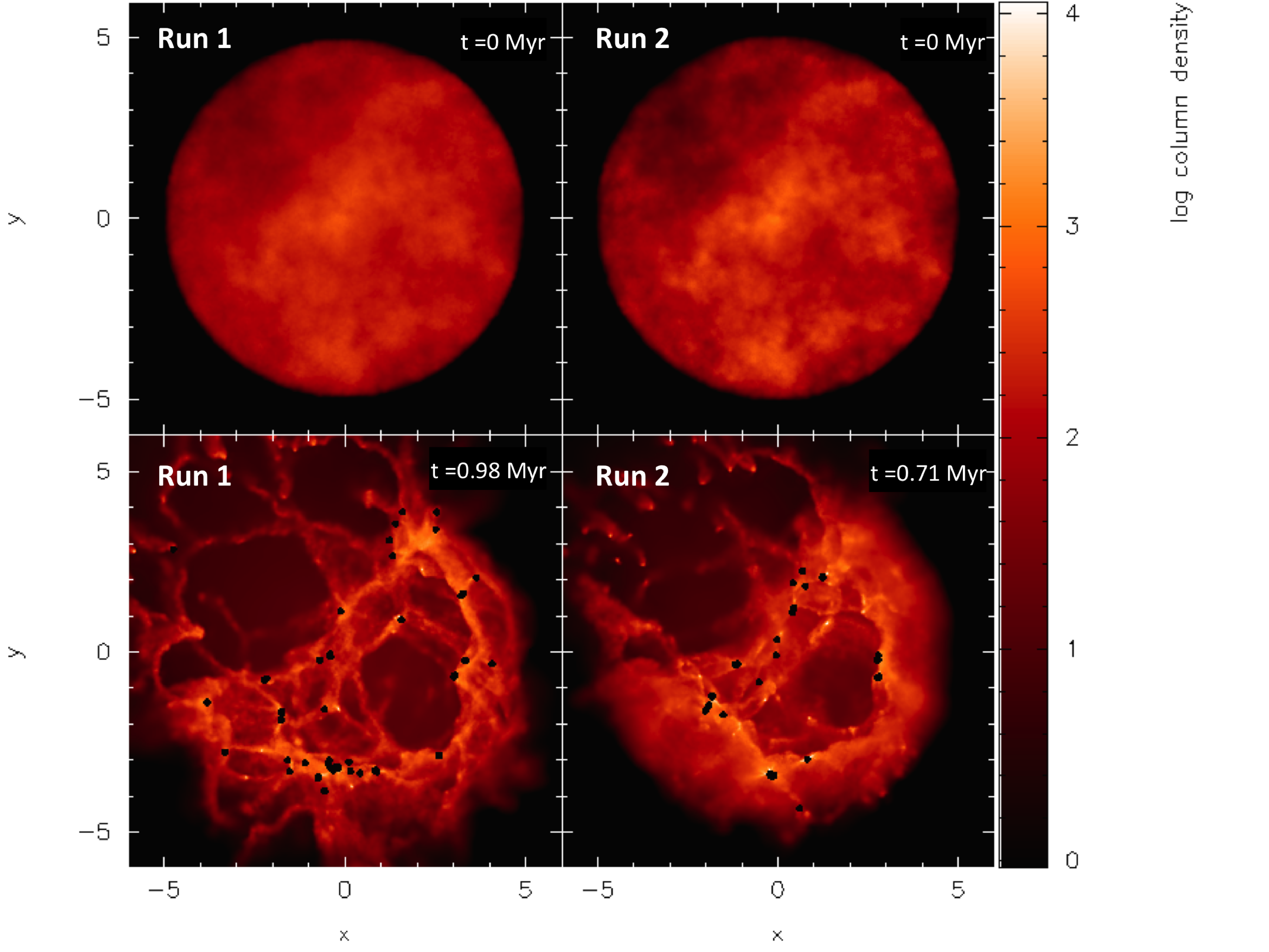}
\caption{{\sc Upper Panels:} Initial column-density distributions in $\,{\rm M}_\odot\,{\rm pc}^{-2}$ for the two runs. In both cases the initial conditions have been generated with the same fractal dimension, $\mathcal{D}=2.4$, and the same random seed, $\mathcal{R}$, but different scaling densities: $\rho_0=1.5$ ({\it Run 1}; left column) and $\rho_0=1.0$ ({\it Run 2}; right column). {\sc Lower Panels:} The column density distributions at $t_{_{\rm END}}$, after $\sim 500\,{\rm M}_\odot$ of the cloud ($\sim 5\%$) has been converted into stars. The black dots mark sink particles, i.e. forming protostars. Each frame is $12\,{\rm pc}\times 12\,{\rm pc}$.}
\label{Figure 2}
\end{figure*}

In this subsection we discuss the results of the SPH simulations. In Figure \ref{Figure 2}, we show the initial and final column density distributions for the simulations. Note that {\it Run 2} has the broader density PDF, and hence the more pronounced density contrasts in the initial conditions. Both setups develop an H{\sc ii} region with a diameter of $\sim 5\,{\rm pc}$. We compare the two simulations at the time when a total mass of $\sim 500\,{\rm M}_\odot$ has been converted into stars, which is $t_{_{\rm END}}=0.98\,{\rm Myr}$ for {\it Run 1} and $t_{_{\rm END}}=0.68\,{\rm Myr}$ for {\it Run 2}. \footnote{Both runs could be followed further, but become extremely time-consuming and slow once the dense shell is collapsing in many places.} The black dots in the evolved H{\sc ii} regions mark sinks, i.e. protostars. At $t_{_{\rm END}}$ there are 79 sinks in {\it Run 1} and 38 sinks in {\it Run 2}. We stress that the simulations are highly dynamical. New protostars are constantly formed and existing protostars continue to accrete at different rates.

\begin{figure}
\includegraphics[width=85mm]{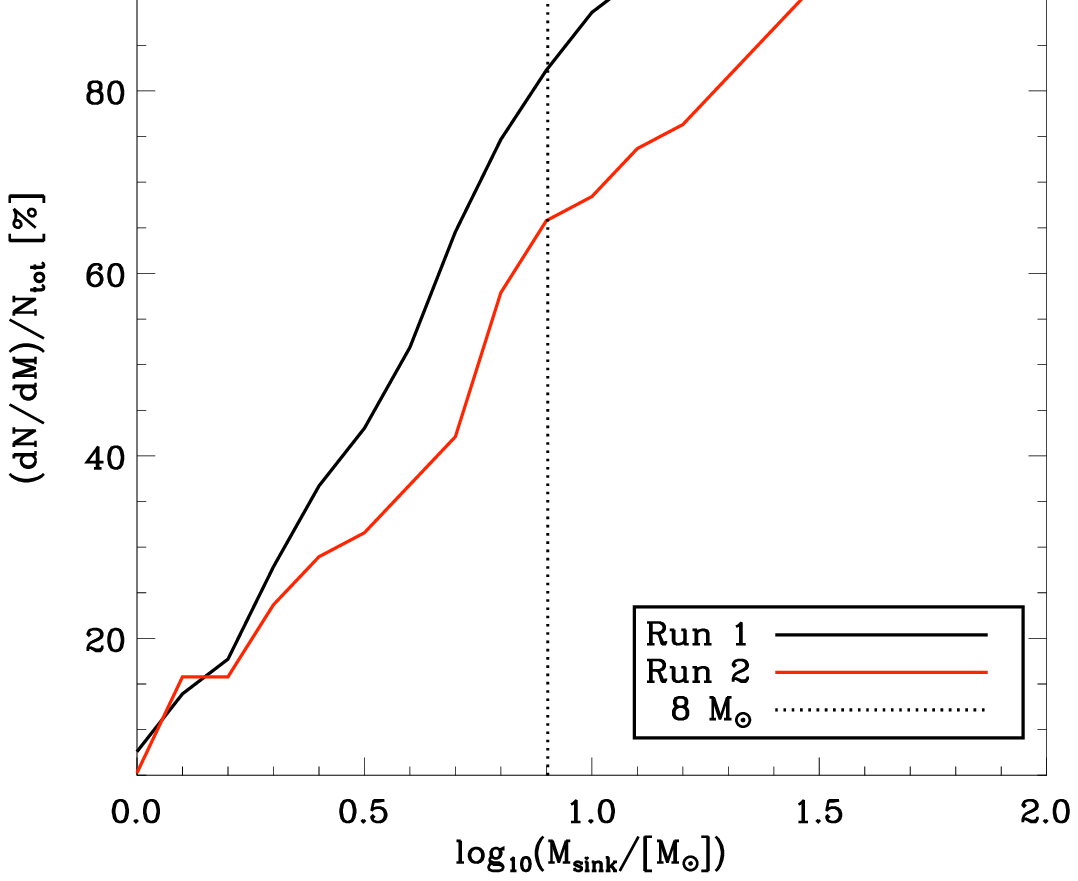}\\
\includegraphics[width=85mm]{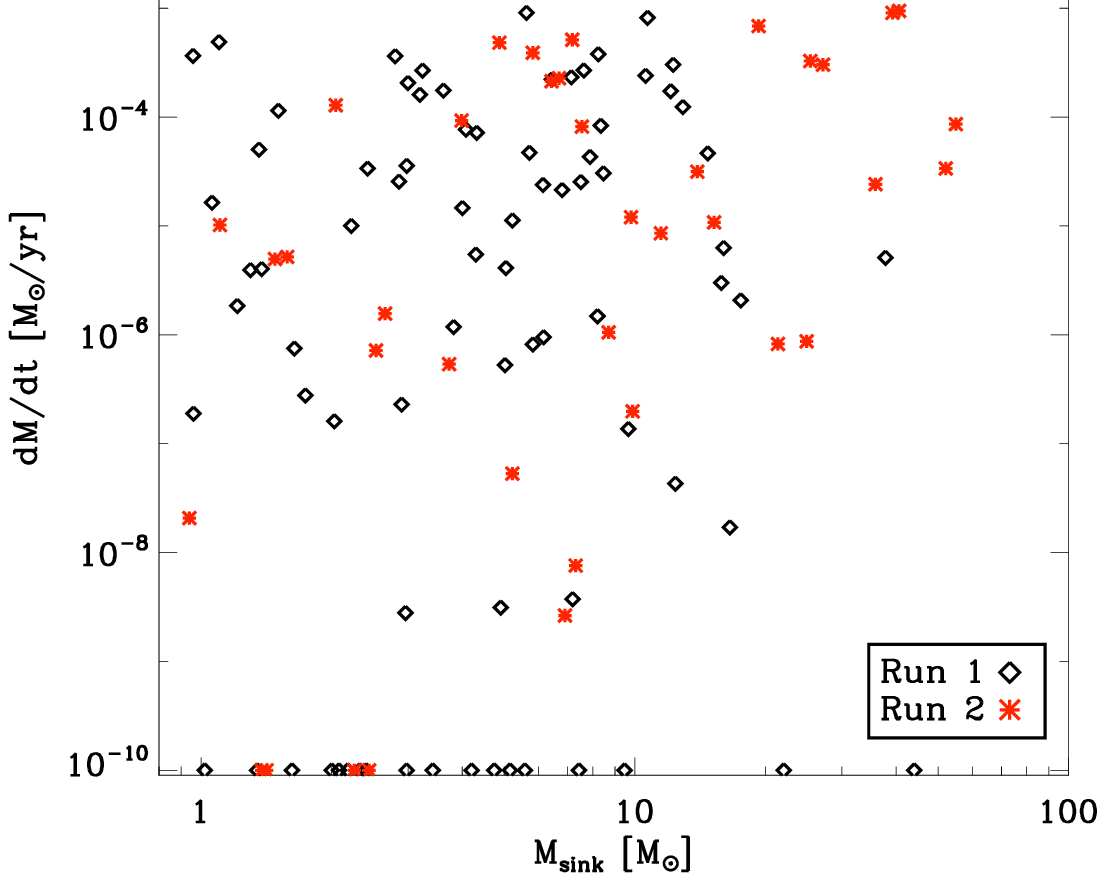}
\caption{{\sc Upper Panel:} Cummulative sink mass functions at $t_{_{\rm END}}$, for {\it Run 1} (black line) and {\it Run 2} (red line), showing that {\it Run 2} forms a higher proportion of massive stars than {\it Run 1}. The vertical dotted line marks $8\,{\rm M}_\odot$, which is usually taken as the lower limit for a massive star, so at this stage $\sim 17\%$ of the stars in {\it Run 1} are massive, and $\sim 34\%$ of the stars in {\it Run 2} are massive {\sc Lower Panel:} Mass accretion rate, $\dot{M}$, plotted against current mass, $M$, at $t_{_{\rm END}}$, for the protostars formed in {\it Run 1} (black open diamonds) and {\it Run 2} (red stars). For protostars that have essentially stopped accreting, we set $\dot{M}=10^{-10}\,{\rm M}_\odot\,{\rm yr}^{-1}$, so they appear in a line along the bottom of the plot; these stars are predominantly formed in {\it Run 1}. For the remaining protostars, there is no strong correlation between $\dot{M}$ and $M$, and no obvious difference between the stars formed in {\it Run 1} and {\it Run 2}, apart from a clutch of eight massive, rapidly accreting stars in {\it Run 2}.}
\label{Figure 3}
\end{figure}

In Figure \ref{Figure 3}, we plot the mass accretion rate onto each protostar as a function of its mass. The masses range from $\sim\!{\rm M}_\odot$ to $\sim\!40\,{\rm M}_\odot$. The accretion rates of protostars that have essentially stopped accreting are arbitrarily set to $\dot{\rm M}_{_{\rm MIN}}=10^{-10}\,{\rm M}_\odot\,{\rm yr}^{-1}$; most of these are from {\it Run 1}. For all other sinks there is no clear correlation between sink mass and mass accretion rate, and the mean accretion rate is $\sim 10^{-5}\,{\rm M}_\odot\,{\rm yr}^{-1}$ in both {\it Runs}. {\it Run 2} forms more massive stars than {\it Run 1}, and has a clutch of eight massive and rapidly accreting protostars at $t_{_{\rm END}}$.

\begin{table}
\begin{center}
\begin{tabular}{ccc|ccc}\hline\hline
\multicolumn{3}{c}{{\it Run 1}} & \multicolumn{3}{c}{\it Run 2}\\\hline 
C & $M_{_{\rm SINK, tot}}$ & $N_{_{\rm SINK}}$ & C & $M_{_{\rm SINK, tot}}$ & $N_{_{\rm SINK}}$\\
 & $\overline{[{\rm M}_\odot]}$ & & & $\overline{[{\rm M}_\odot]}$ & \\\hline 
Shell & 400. & 31 & Shell & 483. & 36 \\
C1 &  0.0  & 0 & C1 &  34.6  & 3 \\
C2 &  0.0  & 0  & C2 &  54.2  & 6 \\
C3 &  140. & 11 & C3 &  171.  & 9 \\\hline\hline
\end{tabular}
\caption{Properties of the protostars embedded in the shell, and in the main clumps. For {\it Run 1} ({\it Run 2}), Column 1 (4) gives the ID of the shell or clump, Column 2 (5) the number of protostars it contains, and Column 3 (6) the total mass of the protostars it contains, at $t_{_{\rm END}}$. Note that the sink masses do not add up to the total mass in protostars at $t_{_{\rm END}}$, because some protostars are located outside the main clumps.}
\label{TAB:msink}
\end{center}
\end{table}

\subsection{Overall bubble structure}

As with RCW 120, we find that the shells formed in our simulations are not perfectly spherical, but elongated and perforated. For example, in the simulations illustrated in Fig. \ref{Figure 2} the initial cloud has a region of reduced column density in the northwest corner, where the ionized gas breaks through and streams out, thereby relieving the pressure of the H{\sc ii} inside the bubble.

The SPH mass distribution does show some small pillars and Evaporating Gaseous Globules (EGGs) close to the northwest hole, but in general pillars are not a prominent feature of H{\sc ii} regions expanding into clouds with fractal dimension $\mathcal{D}=2.4$, because -- for this fractal dimension -- large-scale density fluctuations dominate the cloud structure \citep{Walch2012}.

\subsection{Synthetic observations}\label{radmc3d}

In a post-processing step we map the SPH density distributions at $t_{_{\rm END}}$ onto a three-dimensional grid of $128^3$ $(0.12\,{\rm pc})^3$ cells, using kernel-weighted interpolation. Hence the spatial resolution of the grid at the distance of RCW120, $D=1.34\,{\rm kpc}$, corresponds to the angular resolution of $19.2''$ achieved when observing RCW 120 with APEX-LABOCA at 870 $\mu\rm{m}$ \citep{Deharveng2009}. All the main clumps and condensations are well resolved with $\ga\! 10^4$ SPH particles. Assuming the standard, uniform gas-to-dust ratio of 100, the distribution of SPH particles translates directly into the distribution of dust. The only modification to this rule is that we assume dust to be destroyed at gas temperatures $T >T_{\rm cut}=1200\,{\rm K}$, and therefore the dust density is set to a small offset value in cells above this temperature, which are exclusively those in the H{\sc ii} region; the offset value is $10^{-3}$ times the minimum dust density in the rest of the computational domain. We only take into account silicate dust grains, on the assumption that these dominate the opacity at $870\,\mu{\rm m}$. We use the standard opacity table for this species given by \citet{Draine1984}.

Using the gridded density distribution, we model the transport of continuum radiation against dust opacity using \textsc{Radmc-3D}\footnote{http://www.ita.uni-heidelberg.de/~dullemond/software/radmc-3d} (version 0.25, written by C.~Dullemond; see also \citealt{Peters2010}). In the first step \textsc{Radmc-3D} performs a \textit{thermal} Monte Carlo (MC) radiative transfer simulation to determine the equilibrium dust temperature distribution. The MC implementation is based on the method of \citet{Bjorkman2001}, but includes various improvements, for instance the continuous absorption method of \citet{Lucy1999}. The total luminosity of all sources within the computational domain is distributed amongst $N_{_{\rm PHOT}}=10\times 10^6$ photon packages, where half of them, i.e. $N_{_{\rm SCAT}}=5\times 10^6$ photon packages, are used to compute scattering events. In one set of radiative transfer calculations we only invoke radiation from the central ionizing star; henceforth we refer to these calculations as {\it ionizing source only}. In addition, we perform radiative transfer calculations in which radiation from the newly-formed protostars is also included; henceforth we refer to these calculations as {\it secondary sources included}, and distinguish quantities derived from these calculations with a superscript $\star$. The luminosities of the protostars are estimated on the assumption that they are ZAMS main-sequence stars. This probably results in a significant underestimate of their time-averaged luminosities, but since the luminosities of young protostars are expected to be highly variable, due to infrequent episodic accretion bursts, this seems the safest way to proceed.

\begin{figure*}
\includegraphics[trim = 40mm 0mm 0mm 0mm, clip,width=180mm]{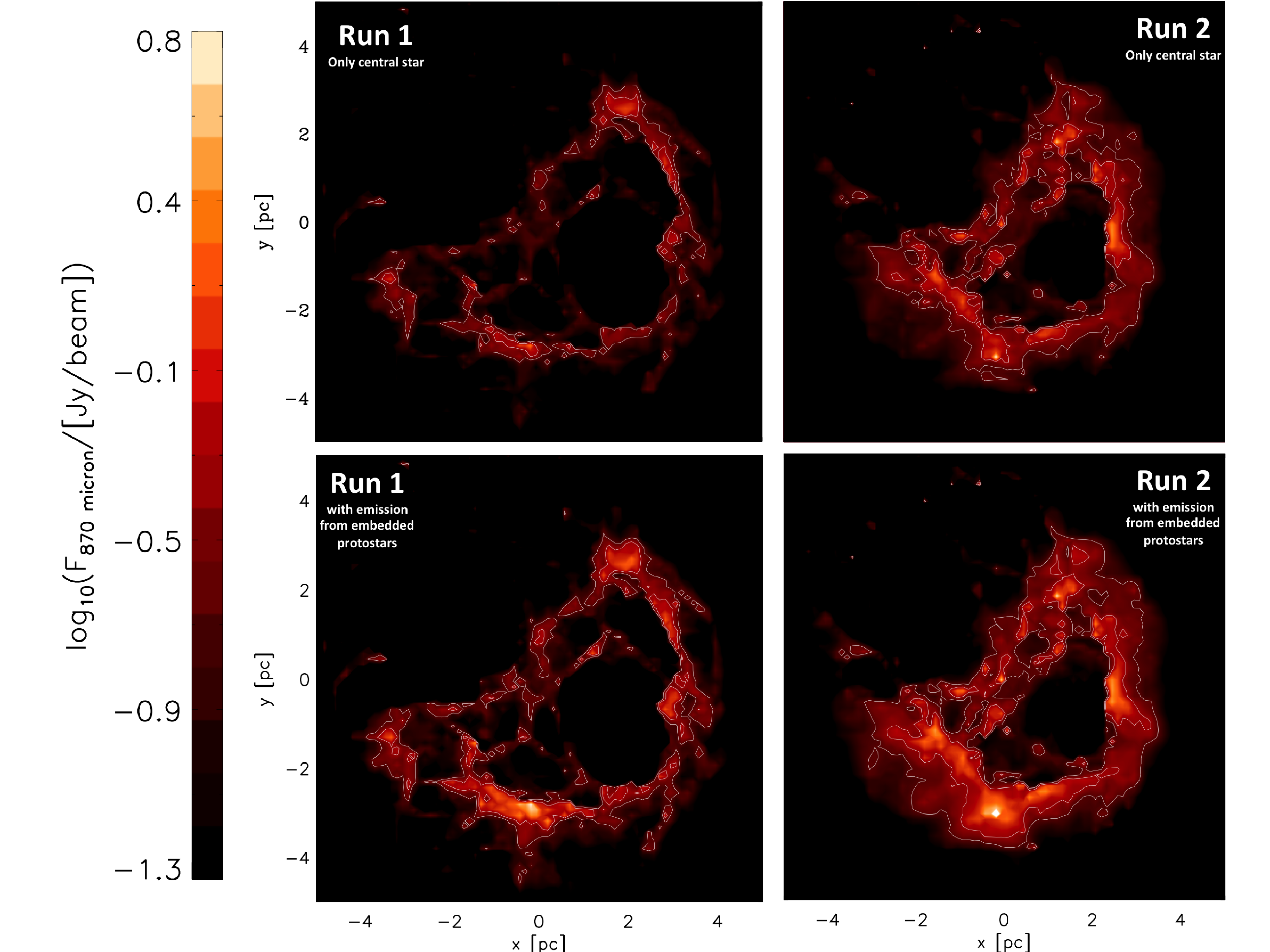}
\caption{870 micron emission calculated using \textsc{RADMC-3D}. The frames on the left refer to {\it Run 1}, and those on the right to {\it Run 2}; the frames on the top row have been calculated with radiation from the ionizing source only, and those on the bottom row with radiation from secondary sources included.The left column shows the images for {\it Run 1}, and the right column for {\it Run 2}.  The white contours are set to fluxes of $0.1\,{\rm Jy}/{\rm beam}$ and $0.5\,{\rm Jy}/{\rm beam}$ in all images. Each frame is $12\,{\rm pc}\times 12\,{\rm pc}$.}
\label{RADMC1}
\end{figure*}

In the second step, \textsc{Radmc-3D} computes isophotal maps at $870\,\mu{\rm m}$, using ray tracing. Figure \ref{RADMC1} shows the resulting isophotal maps of our simulations, as seen in their ($x,y$)-projection. The outer white contour marks the $0.1\,{\rm Jy}/\rm{beam}$ cutoff value, which we use to define the total mass of the shell. The inner white contour marks the $0.5\,{\rm Jy}/\rm{beam}$ value, which we use to define the masses of the clumps. With radiation from the ionizing source only (top row of Fig. \ref{RADMC1}), the total flux an observer at $1.34\,{\rm kpc}$ would receive at $870\,\mu{\rm m}$ is $F_{_{\rm 870}}^{\rm tot}=312\,{\rm Jy}$ for {\it Run 1} and $F_{_{\rm 870}}^{\rm tot}=535\,{\rm Jy}$ for {\it Run 2}. If radiation from the secondary sources is included (bottom row of Fig. \ref{RADMC1}) the resulting total fluxes are significantly higher: $F_{_{\rm 870}}^{\rm tot}=500\,{\rm Jy}$ for {\it Run 1} and $F_{_{\rm 870}}^{\rm tot}=760\,{\rm Jy}$ for {\it Run 2}. The fluxes are higher in the second case, because most of the newly formed protostars are located in the shell, and in or near the clumps.

Qualitatively, the synthetic isophotal maps of our simulations are very similar to the $870\,\mu{\rm m}$ observations of RCW 120. In the following, we compare synthetic and observed images in greater detail, in order to assess the SPH simulation, but also to evaluate the uncertainty of the mass estimates for clumps in the shell of RCW 120.

We use the $0.5\,{\rm Jy}/\rm{beam}$ contour to define clumps within the shell. In particular, we identify the 3 most massive clumps, which we label C1, C2, C3, and focus our analysis on their properties. The total mass and number of sinks located within each clump are listed in Table \ref{TAB:msink}. The mass in gas and dust in the shells and in the individual clumps are listed in Table 2, which we describe more thoroughly in the next section. 
\section{Discussion}\label{sec4}

\begin{figure*}
\includegraphics[trim = 40mm 0mm 0mm 0mm, clip, width=180mm]{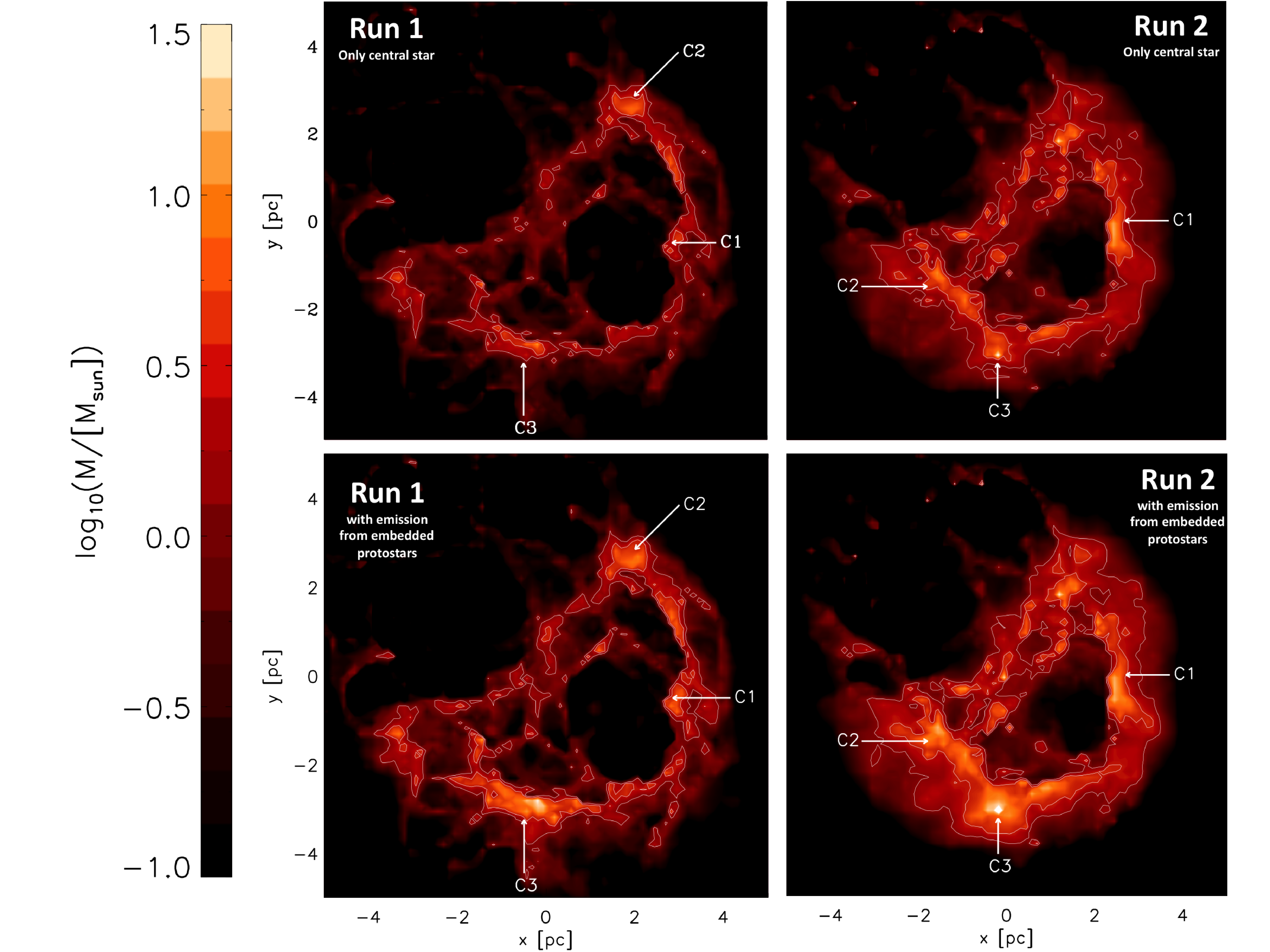}
\caption{Mass distribution derived from the thermal dust emission at $870\, \mu{\rm m}$ shown in Fig. \ref{RADMC1}. For comparison we overlay contours marking $0.1\,{\rm Jy}/{\rm beam}$ and $0.5\,{\rm Jy}/{\rm beam}$ from Fig. \ref{RADMC1}. As in Fig. \ref{RADMC1}, the frames on the left refer to {\it Run 1}, and those on the right to {\it Run 2}; the frames on the top row have been calculated with radiation from the ionizing source only, and those on the bottom row with radiation from secondary sources included. The three largest clumps, which are discussed in the text, are marked with C1, C2, and C3. Each frame is $12\,{\rm pc}\times 12\,{\rm pc}$.}
\label{MASS1}
\end{figure*}

\subsection{Clump masses}\label{clumpmass}

In order to compare the simulations with observations of RCW 120, we adopt the same procedure as in \citet{Deharveng2009} to calculate the masses of the shells, and of the individual clumps, from the synthetic emission maps,
\begin{equation}\label{mdust}
M_{_{870}}=100\,\frac{F_{_{870}}\,D^2}{\kappa_{_{870}}\,B_{_{870}}(T_{_{\rm{DUST}}})}\,.
\end{equation}
Here $D$ is the distance of the source ($D=1.34 \,\rm{kpc}$ for RCW 120), $\kappa_{_{870}}=1.8\,\rm{cm}^2\,\rm{g}^{-1}$ \citep{Ossenkopf1994} is the opacity per unit mass of dust at $870\,\mu{\rm m}$, and $B_{_{870}}(T_{_{\rm{DUST}}})$ is the Planck function at 870 micron for dust temperature $T_{_{\rm{DUST}}}$. We adopt $T_{_{\rm{DUST}}}=30\,\rm{K}$, since we do not allow the gas to cool below this temperature in the simulations; \citet{Deharveng2009} also invoke this temperature to analyse their results. A constant gas-to-dust mass ratio of 100 has been assumed. 

In Figure \ref{MASS1} we show the mass distribution derived from the isophotal maps using Eq. \ref{mdust}. For reference we overplot the $0.1\,{\rm Jy}/\rm{beam}$ and $0.5 \; {\rm Jy}/\rm{beam}$ contours from Fig. \ref{RADMC1}, which define -- respectively -- the shell and the most massive clumps. For {\it Run 1} we calculate a total shell mass of $1818\,{\rm M}_\odot$ with radiation from the ionizing source only, and $2928\,{\rm M}_\odot$ with radiation from secondary sources included. Both of these estimates are similar to the shell mass of $1100\,{\rm M}_\odot$ found in RCW 120 with $T_{_{\rm DUST}}=30\,{\rm K}$. For {\it Run 2} the corresponding shell masses are $3013\,{\rm M}_\odot$ and $4420\,{\rm M}_\odot$.

\subsection{Massive clumps without C\&C}

We divide the shells up into clumps using the $0.5\,{\rm Jy}/\rm{beam}$ contour on the synthetic isophotal maps. These clumps typically have masses between a few and a few hundred ${\rm M}_\odot$. In Fig. \ref{MASS1} we identify the three main clumps formed in each {\it Run}, and their properties are listed in Table \ref{TAB:clumpmass}. All these clumps are sufficiently massive that they may spawn massive protostars, leading to sequential propagation of massive star formation \citep[see][ for observational evidence of sequential triggering]{Koenig2012}.

The individual clumps are well aligned with the overall structure of the ionisation front and the swept-up dense shell. However, our simulations clearly show that the formation of separate clumps in these simulations is not really due to C\&C -- in the sense that at no time do we observe the formation of a coherent shell, which grows to subsequently become gravitationally unstable and undergo fragmentation. The seeds for the clumps are already present in the initial fractal density structure of the cloud, and as the H{\sc ii} region expands it sweeps additional low-density material into the clumps, and the clumps themselves are pushed outwards and collect additional material that way. At the same time, the ionizing radiation penetrates low-density regions much more easily than high-density regions, and as a result the hot, high-pressure ionised gas tends to envelop the dense clumps and compress them, as in Radiatively Driven Implosion. It is the combination of collecting additional mass {\it and} being enveloped by the H{\sc ii} region that renders the clumps unstable and drives them into collapse. This is therefore a hybrid process, combining elements of both C\&C and RDI. Based on 2D radiation-hydrodynamic simulations, \citet{Elmegreen1995} discuss a similar scenario for massive core formation by shock focusing of turbulent clumps inside a moving post-shock layer.

We are unable to comment on whether individual protostars are regularly spaced within individual clumps (as noted by \citet{Deharveng2009} in the unsharp-masked $24\,\mu{\rm m}$ Spitzer image of their Condensation 1) because properly modelling the $24\,\mu{\rm m}$ emission is beyond the scope of this paper \citep[see, for example, ][]{Koepferl2015}. However, we note that regularly spaced protostars could also arise in the scenario we have described above; they are not necessarily a product of C\&C.\\
The masses estimated for the three main clumps are strongly dependent on whether the radiative transfer modeling includes the radiation from secondary sources or not (see Table \ref{TAB:clumpmass} and section \ref{SEC:Massestimate}). The variations can be as high as a factor of 8, as is the case for clump C3 in {\it Run 1}, for which the estimated mass increases from $61\,{\rm M}_\odot$ to $493\,{\rm M}_\odot$ when the radiation from secondary sources is included. This is because the extra heating from newly-formed protostars makes the dust in their vicinity hotter, and therefore more material falls within the  $0.5\,{\rm Jy}/\rm{beam}$ threshold. 

\subsection{How reliable is the mass distribution obtained from $870\,\mu{\rm m}$ fluxes?}\label{SEC:Massestimate}

In this section we compare how well the actual mass distribution in the simulations is recovered by applying Eq. \ref{mdust} to the synthetic isophotal maps at $870\, \mu{\rm m}$. This comparison can provide useful insights into the reliability of clump mass estimates from observational data.

We define four masses, and list their values in Table \ref{TAB:clumpmass}. $M_{_{870}}$ is the mass obtained using Eq. \ref{mdust} on synthetic isophotal maps calculated with radiation from the ionizing source only, and $M_{_{870}}^{\star}$ is the mass obtained in the same way, but with secondary sources included. Likewise, $M_{_{\rm TRUE}}$ is the actual mass falling within a shell or clump on synthetic $870\, \mu{\rm m}$ isophotal maps calculated with radiation from the ionizing source only, whilst $M_{_{\rm TRUE}}^{\star}$ is the corresponding quantity obtained when radiation from secondary sources is included in the radiation transfer modelling. $M_{_{\rm TRUE}}$ and $M_{_{\rm TRUE}}^{\star}$ are obtained by integrating the surface density of SPH particles over the area covered by the shell or clump (i.e. the area inside the $0.1\,{\rm Jy}/{\rm beam}$ or $0.5\,{\rm Jy}/{\rm beam}$ contours, respectively).

\begin{figure}
\includegraphics[trim = 10mm 0mm 0mm 0mm, clip,width=85mm]{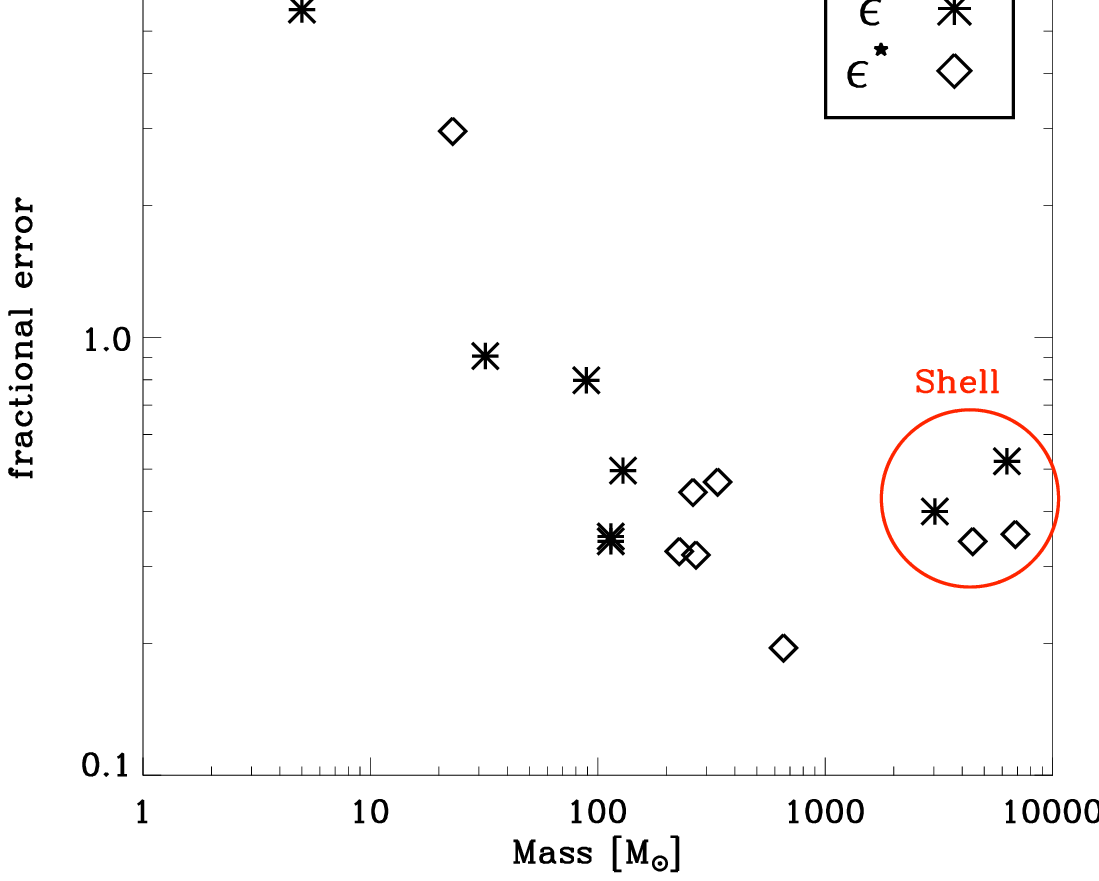}
\caption{Fractional errors in the masses of shells and clumps obtained using Eq. \ref{mdust} on synthetic isophotal maps, with and without radiation from newly-formed protostars (see Eqs. \ref{EQN:epsilon} and \ref{EQN:epsilonstar}), plotted against their true masses; a value of 1 is a perfect match. Open diamonds represent results from {\it Run 1}, and stars represent results from {\it Run 2}. Values pertaining to the shell are contained within the red ring; all the other values represent clumps.}
\label{Fig_merr}
\end{figure}

The resulting fractional  errors, 
\begin{eqnarray}\label{EQN:epsilon}
\epsilon&=&\frac{\left| M_{_{870}} - M_{_{\rm TRUE}}\right|}{M_{_{\rm TRUE}}}\,,
\end{eqnarray}
and 
\begin{eqnarray}\label{EQN:epsilonstar}
\epsilon^\star&=&\frac{\left| M_{_{870}}^\star - M_{_{\rm TRUE}}^\star\right|}{M_{_{\rm TRUE}}^\star}\,,
\end{eqnarray}
are plotted against the true masses ($M_{_{\rm TRUE}}$, $M_{_{\rm TRUE}}^\star$) on Fig. \ref{Fig_merr}. The {\it shell} masses obtained using Eq. \ref{mdust} on synthetic $870\;\mu{\rm m}$ isophotal maps are always lower than the true masses, typically by a factor $\la\!2$, irrespective of whether the radiation from secondary sources is included or not. In contrast, the masses of individual {\it clumps} obtained using Eq. \ref{mdust} on synthetic $870\;\mu{\rm m}$ isophotal maps are sometimes higher and sometimes lower than the true masses; for massive clumps ($>100\,{\rm M}_\odot$) the fractional error is always less than two, but for lower-mass clumps it can be very large, and such masses should probably not be trusted. 

\begin{figure}
\includegraphics[trim = 27mm 0mm 0mm 0mm, clip,width=90mm]{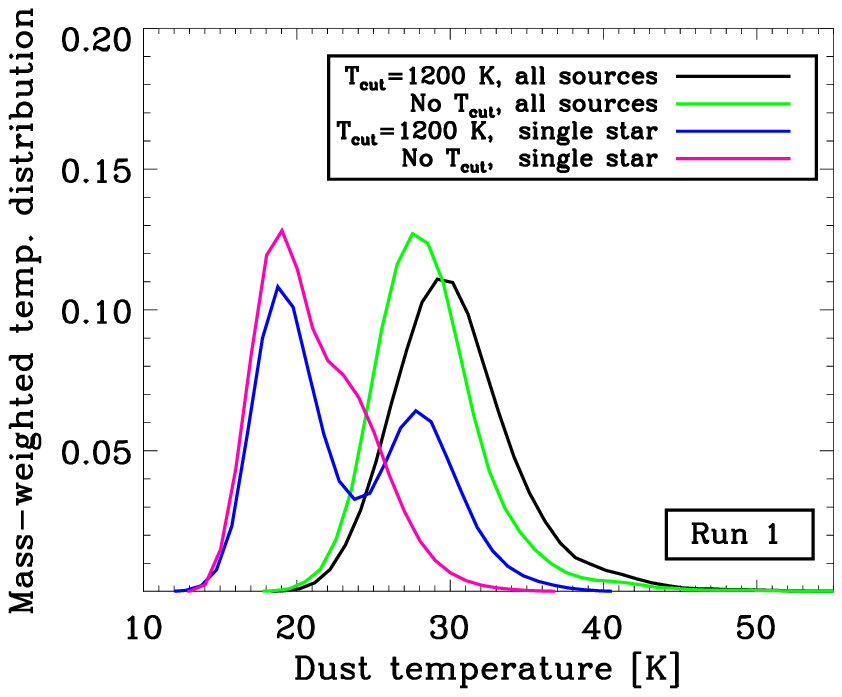}
\caption{Mass-weighted dust temperature distribution for different {\sc Radmc}-3D models of {\it Run 1}. All dust temperature distributions were derived using $10^7$ photon packages. The black line shows the fiducial case, where we include the emission from triggered protostars and cut the dust abundance at gas temperatures above $T_\mathrm{cut}=1200$ K, while the green line ('No $T_\mathrm{cut}$') shows the case where also dust within the H{\sc ii} region is included. Most of the mass is found at $T_\mathrm{dust}=30$ K. In the latter case the distribution shifts to a slightly colder mean temperature since less photons are available to heat the dust within the shell due to the increased absorption within the H{\sc ii} region. The blue and pink lines show the resulting dust temperature distribution in the case that only the emission from the central O-star is taken into account. 
\label{FIG:TempHistogram}}
\end{figure}

In general, masses obtained using Eq. \ref{mdust} on synthetic $870\;\mu{\rm m}$ maps calculated including radiation from secondary sources are more accurate than those obtained with radiation from the ionising source only, as they should be. In order to quantify the effect of including secondary sources, we have generated the mass-weighted distribution of temperature, both with and without secondary sources; in addition, we have generated the corresponding distributions for radiation transport calculations without the gas-temperature cut-off at $T_{\rm cut}=1200\,{\rm K}$, so that emission from dust within the HII region is included. The resulting distributions for {\it Run 1} are shown in Fig. \ref{FIG:TempHistogram}; similar results are obtained for {\it Run 2}. We note two features of this plot. First, when there is no cut-off, the mean dust temperature is lower. This is because the dust in the H{\sc ii} region absorbs much of the short wavelength radiation from the ionising source and the newly-formed protostars, and then radiates it at long wavelengths, with the result that it escapes without heating the neutral gas further out. Second, when the secondary sources are included, the mean dust-temperature is almost exactly $30\,{\rm K}$, as we have assumed following the observational interpretation of RCW 120.

\begin{table}
\begin{center}
\begin{tabular}{ccccc}
\hline\hline
ID & $M_{_{870}}$ & $M_{_{870}}^{\star}$ & $M_{_{\rm TRUE}}$ & $M_{_{\rm TRUE}}^{\star}$\\
  & $\overline{{\rm M}_\odot}$ & $\overline{{\rm M}_\odot}$ & $\overline{{\rm M}_\odot}$ & $\overline{{\rm M}_\odot}$  \\\hline
\multicolumn{5}{c}{\it Run 1}\\\hline 
Shell & 1818 & 2928 & 3031 & 4450 \\
C1    & 33   & 91   & 5.0   &  23.\\
C2    & 75   & 146  & 114  &  262 \\ 
C3    & 61   & 493  & 32.  &  336 \\\hline
\multicolumn{5}{c}{\it Run 2}\\\hline
Shell & 3013 & 4420 & 6291 & 6853 \\
C1    & 193  & 302  & 129 & 228 \\
C2    & 74   & 184  & 114 & 270  \\ 
C3    & 160  & 783  & 89 & 655  \\\hline
\end{tabular}
\caption{Column 1 identifies the different structures analysed. Columns 2 and 3 give mass estimates derived by applying Eq. 5 to the synthetic $870\;\mu{\rm m}$ images which were calculated with radiation from the ionizing source only, $M_{_{870}}$, and with radiation from secondary sources included, $M_{_{870}}^{\star} $ (see Fig. 5). Columns 4 and 5 give the true masses, obtained by integrating the SPH column density over the area inside the $0.1 \;{\rm Jy}/{\rm beam}$ contour for shells, and the $0.5 \;{\rm Jy}/{\rm beam}$ contour for the clumps. $M_{_{\rm TRUE}} $ is the mass within these contour levels calculated with radiation from the ionizing star only, whereas $M_{_{\rm TRUE}}^{\star}$ is the mass within these contour levels calculated with radiation from the secondary sources included. Both, $M_{_{\rm TRUE}}$ and $M_{_{\rm TRUE}}^{\star}$ account only for the mass of gas and dust and do not include embedded protostars; the total mass of embedded protostars is listed in Table 1.}
\label{TAB:clumpmass}
\end{center}
\end{table}

\section{Conclusions}\label{Conclusions}

We have performed high resolution SPH simulations of H{\sc ii} regions expanding into fractal molecular clouds, and compared synthetic $870\,\mu{\rm m}$ isophotal maps of these simulations (obtained using \textsc{Radmc-3D}) to $870\,\mu{\rm m}$ observations of the well-studied, galactic H{\sc ii} region RCW 120. Our model reproduces a swept-up shell with a mass between 3000 and $7000\,{\rm M}_\odot$. The shell contains several star-forming clumps, with masses between 30 and $700\,{\rm M}_\odot$. These cores are not formed by shell fragmentation.

We suggest that finding massive clumps and sites of high mass star formation within an expanding shell around an H{\sc ii} region need not, and probably should not, be taken as evidence for the Collect \& Collapse mechanism at work. The clumpiness of the dense shell can be attributed to density structures in the fractal, turbulent, molecular cloud into which the H{\sc ii} regon expands. Two processes combine to render the larger clumps gravitationally unstable. First the expanding H{\sc ii} region sweeps lower-density gas into the clumps, and pushes the clumps outwards so they also sweep up lower-density gas further out. Second, the H{\sc ii} region advances more rapidly through the lower-density gas around a clump, thereby enveloping and squeezing it. In other words, a hybrid mechanism is at work, which combines elements of C\&C and RDI; it is not standard C\&C because at no stage does a coherent shell form and then become gravitationally unstable and fragment, but it relates to C\&C because the clumps do collect additional material, due to the expansion of the H{\sc ii} region. A detailed study of the growth of clumps in the shells of expanding H{\sc ii} regions is presented in \citet{Walch2013}.\\
Overall, we find good agreement (to within a factor of 2) between the actual mass distribution and the mass distribution inferred from thermal dust emission, for shells and clumps having masses greater than $\,\sim\! 100\,{\rm M}_\odot$. In particular, masses estimated from synthetic $870\,\mu{\rm m}$ isophotal maps always under-predict the mass of the the large-scale {\it shell} structure, whereas for individual {\it clumps} both over- and under-estimates are possible, and usually the result is accurate to within a factor of two. To obtain a realistic estimate of the shell and clump masses from synthetic $870\,\mu{\rm m}$ isophotal maps, it is important to take into account the heating from embedded protostars.



\section*{Acknowledgments}
We thank the anonymous referees for constructive comments on the manuscript.
SW and AW acknowledge the support of the Marie Curie RTN \textsc{CONSTELLATION}.
SW further thanks the German Science Foundation for their support in the framework of the ISM priority programme 1573 and through SFB 956 on 'The conditions and impact of star formation'.
The work of TGB was funded by STFC grant ST/J001511/1.
RW acknowledges support by the Czech Science Foundation grant 209/12/1795 and by the project RVO: 67985815; the Academy of
Sciences of the Czech Republic. 
The column density plots were made using the SPLASH visualization code \citep{Price2007}.

\bibliographystyle{mn2e}
\bibliography{references}

\begin{thebibliography}{}

\bibitem[\protect\citeauthoryear{{Anderson}, {Zavagno}, {Deharveng}, {Abergel},
  {Motte}, {Andr{\'e}}, {Bernard}, {Bontemps}, {Hennemann}, {Hill},
  {Rod{\'o}n}, {Roussel} \& {Russeil}}{{Anderson} et~al.}{2012}]{Anderson2012}
{Anderson} L.~D.,  {Zavagno} A.,  {Deharveng} L.,  {Abergel} A.,  {Motte} F.,
  {Andr{\'e}} P.,  {Bernard} J.-P.,  {Bontemps} S.,  {Hennemann} M.,  {Hill}
  T.,  {Rod{\'o}n} J.~A.,  {Roussel} H.,    {Russeil} D.,  2012, \aap, 542, A10

\bibitem[\protect\citeauthoryear{{Balsara}}{{Balsara}}{1995}]{Balsara1995}
{Balsara} D.~S.,  1995, Journal of Computational Physics, 121, 357

\bibitem[\protect\citeauthoryear{{Barnes} \& {Hut}}{{Barnes} \&
  {Hut}}{1986}]{Barnes1986}
{Barnes} J.,  {Hut} P.,  1986, \nat, 324, 446

\bibitem[\protect\citeauthoryear{{Bate} \& {Burkert}}{{Bate} \&
  {Burkert}}{1997}]{BB1997}
{Bate} M.~R.,  {Burkert} A.,  1997, \mnras, 288, 1060

\bibitem[\protect\citeauthoryear{{Bisbas}, {W{\"u}nsch}, {Whitworth} \&
  {Hubber}}{{Bisbas} et~al.}{2009}]{Bisbas2009}
{Bisbas} T.~G.,  {W{\"u}nsch} R.,  {Whitworth} A.~P.,    {Hubber} D.~A.,  2009,
  \aap, 497, 649

\bibitem[\protect\citeauthoryear{{Bisbas}, {W{\"u}nsch}, {Whitworth}, {Hubber}
  \& {Walch}}{{Bisbas} et~al.}{2011}]{Bisbas2011}
{Bisbas} T.~G.,  {W{\"u}nsch} R.,  {Whitworth} A.~P.,  {Hubber} D.~A.,
  {Walch} S.,  2011, \apj, 736, 142

\bibitem[\protect\citeauthoryear{{Bjorkman} \& {Wood}}{{Bjorkman} \&
  {Wood}}{2001}]{Bjorkman2001}
{Bjorkman} J.~E.,  {Wood} K.,  2001, \apj, 554, 615

\bibitem[\protect\citeauthoryear{{Churchwell}, {Povich}, {Allen}, {Taylor},
  {Meade}, {Babler}, {Indebetouw}, {Watson}, {Whitney}, {Wolfire}, {Bania},
  {Benjamin}, {Clemens}, {Cohen} \& {Cyganowski}}{{Churchwell}
  et~al.}{2006}]{Churchwell2006}
{Churchwell} E.,  {Povich} M.~S.,  {Allen} D.,  {Taylor} M.~G.,  {Meade} M.~R.,
   {Babler} B.~L.,  {Indebetouw} R.,  {Watson} C.,  {Whitney} B.~A.,  {Wolfire}
  M.~G.,  {Bania} T.~M.,  {Benjamin} R.~A.,  {Clemens} D.~P.,  {Cohen} M.,
  {Cyganowski} C.~J.,  2006, \apj, 649, 759

\bibitem[\protect\citeauthoryear{{Dale}, {Bonnell} \& {Whitworth}}{{Dale}
  et~al.}{2007}]{Dale2007}
{Dale} J.~E.,  {Bonnell} I.~A.,    {Whitworth} A.~P.,  2007, \mnras, 375, 1291

\bibitem[\protect\citeauthoryear{{Dale}, {W{\"u}nsch}, {Whitworth} \& {Palou{\v
  s}}}{{Dale} et~al.}{2009}]{Dale2009}
{Dale} J.~E.,  {W{\"u}nsch} R.,  {Whitworth} A.,    {Palou{\v s}} J.,  2009,
  \mnras, 398, 1537

\bibitem[\protect\citeauthoryear{{Deharveng}, {Lefloch}, {Zavagno}, {Caplan},
  {Whitworth}, {Nadeau} \& {Mart{\'{\i}}n}}{{Deharveng}
  et~al.}{2003}]{Deharveng2003}
{Deharveng} L.,  {Lefloch} B.,  {Zavagno} A.,  {Caplan} J.,  {Whitworth} A.~P.,
   {Nadeau} D.,    {Mart{\'{\i}}n} S.,  2003, \aap, 408, L25

\bibitem[\protect\citeauthoryear{{Deharveng}, {Zavagno}, {Schuller}, {Caplan},
  {Pomar{\`e}s} \& {De Breuck}}{{Deharveng} et~al.}{2009}]{Deharveng2009}
{Deharveng} L.,  {Zavagno} A.,  {Schuller} F.,  {Caplan} J.,  {Pomar{\`e}s} M.,
     {De Breuck} C.,  2009, \aap, 496, 177

\bibitem[\protect\citeauthoryear{{Draine} \& {Lee}}{{Draine} \&
  {Lee}}{1984}]{Draine1984}
{Draine} B.~T.,  {Lee} H.~M.,  1984, \apj, 285, 89

\bibitem[\protect\citeauthoryear{{Elmegreen}}{{Elmegreen}}{1994}]{Elmegreen199%
4}
{Elmegreen} B.~G.,  1994, \apj, 427, 384

\bibitem[\protect\citeauthoryear{{Elmegreen}, {Kimura} \& {Tosa}}{{Elmegreen}
  et~al.}{1995}]{Elmegreen1995}
{Elmegreen} B.~G.,  {Kimura} T.,    {Tosa} M.,  1995, \apj, 451, 675

\bibitem[\protect\citeauthoryear{{Elmegreen} \& {Lada}}{{Elmegreen} \&
  {Lada}}{1977}]{Elmegreen1977}
{Elmegreen} B.~G.,  {Lada} C.~J.,  1977, \apj, 214, 725

\bibitem[\protect\citeauthoryear{{Falgarone}, {Phillips} \&
  {Walker}}{{Falgarone} et~al.}{1991}]{Falgarone1991}
{Falgarone} E.,  {Phillips} T.~G.,    {Walker} C.~K.,  1991, \apj, 378, 186

\bibitem[\protect\citeauthoryear{{Federrath}, {Klessen} \&
  {Schmidt}}{{Federrath} et~al.}{2008}]{Federrath2008}
{Federrath} C.,  {Klessen} R.~S.,    {Schmidt} W.,  2008, \apjl, 688, L79

\bibitem[\protect\citeauthoryear{{Federrath}, {Klessen} \&
  {Schmidt}}{{Federrath} et~al.}{2009}]{Federrath2009}
{Federrath} C.,  {Klessen} R.~S.,    {Schmidt} W.,  2009, \apj, 692, 364

\bibitem[\protect\citeauthoryear{{Gritschneder}, {Burkert}, {Naab} \&
  {Walch}}{{Gritschneder} et~al.}{2010}]{Gritschneder2010}
{Gritschneder} M.,  {Burkert} A.,  {Naab} T.,    {Walch} S.,  2010, \apj, 723,
  971

\bibitem[\protect\citeauthoryear{{Gritschneder}, {Naab}, {Walch}, {Burkert} \&
  {Heitsch}}{{Gritschneder} et~al.}{2009}]{Gritschneder2009}
{Gritschneder} M.,  {Naab} T.,  {Walch} S.,  {Burkert} A.,    {Heitsch} F.,
  2009, \apjl, 694, L26

\bibitem[\protect\citeauthoryear{{Haworth} \& {Harries}}{{Haworth} \&
  {Harries}}{2012}]{Haworth2012}
{Haworth} T.~J.,  {Harries} T.~J.,  2012, \mnras, 420, 562

\bibitem[\protect\citeauthoryear{{Hubber}, {Batty}, {McLeod} \&
  {Whitworth}}{{Hubber} et~al.}{2011}]{Hubber2011}
{Hubber} D.~A.,  {Batty} C.~P.,  {McLeod} A.,    {Whitworth} A.~P.,  2011,
  \aap, 529, A27+

\bibitem[\protect\citeauthoryear{{Hubber}, {Walch} \& {Whitworth}}{{Hubber}
  et~al.}{2013}]{Hubber2013}
{Hubber} D.~A.,  {Walch} S.,    {Whitworth} A.~P.,  2013, \mnras, 430, 3261

\bibitem[\protect\citeauthoryear{{Kessel-Deynet} \& {Burkert}}{{Kessel-Deynet}
  \& {Burkert}}{2003}]{Kessel2003}
{Kessel-Deynet} O.,  {Burkert} A.,  2003, \mnras, 338, 545

\bibitem[\protect\citeauthoryear{{Koenig}, {Leisawitz}, {Benford}, {Rebull},
  {Padgett} \& {Assef}}{{Koenig} et~al.}{2012}]{Koenig2012}
{Koenig} X.~P.,  {Leisawitz} D.~T.,  {Benford} D.~J.,  {Rebull} L.~M.,
  {Padgett} D.~L.,    {Assef} R.~J.,  2012, \apj, 744, 130

\bibitem[\protect\citeauthoryear{{Koepferl}, {Robitaille}, {Morales} \&
  {Johnston}}{{Koepferl} et~al.}{2015}]{Koepferl2015}
{Koepferl} C.~M.,  {Robitaille} T.~P.,  {Morales} E.~F.~E.,    {Johnston}
  K.~G.,  2015, \apj, 799, 53

\bibitem[\protect\citeauthoryear{{Lee}}{{Lee}}{2004}]{Lee2004}
{Lee} Y.,  2004, Journal of Korean Astronomical Society, 37, 137

\bibitem[\protect\citeauthoryear{{Lucy}}{{Lucy}}{1999}]{Lucy1999}
{Lucy} L.~B.,  1999, \aap, 344, 282

\bibitem[\protect\citeauthoryear{{Masunaga}, {Miyama} \& {Inutsuka}}{{Masunaga}
  et~al.}{1998}]{Masunaga1998}
{Masunaga} H.,  {Miyama} S.~M.,    {Inutsuka} S.-i.,  1998, \apj, 495, 346

\bibitem[\protect\citeauthoryear{{Monaghan}}{{Monaghan}}{1992}]{Monaghan1992}
{Monaghan} J.~J.,  1992, \araa, 30, 543

\bibitem[\protect\citeauthoryear{{Monaghan} \& {Gingold}}{{Monaghan} \&
  {Gingold}}{1983}]{Monaghan1983}
{Monaghan} J.~J.,  {Gingold} R.~A.,  1983, Journal of Computational Physics,
  52, 374

\bibitem[\protect\citeauthoryear{{Ossenkopf} \& {Henning}}{{Ossenkopf} \&
  {Henning}}{1994}]{Ossenkopf1994}
{Ossenkopf} V.,  {Henning} T.,  1994, \aap, 291, 943

\bibitem[\protect\citeauthoryear{{Osterbrock}}{{Osterbrock}}{1974}]{Osterbrock%
1974}
{Osterbrock} D.~E.,  1974, {Astrophysics of gaseous nebulae}

\bibitem[\protect\citeauthoryear{{Osterbrock} \& {Ferland}}{{Osterbrock} \&
  {Ferland}}{2006}]{Osterbrock2006}
{Osterbrock} D.~E.,  {Ferland} G.~J.,  2006, {Astrophysics of gaseous nebulae
  and active galactic nuclei}.
{University Science Books}

\bibitem[\protect\citeauthoryear{{Padoan}, {Jones} \& {Nordlund}}{{Padoan}
  et~al.}{1997}]{Padoan1997}
{Padoan} P.,  {Jones} B.~J.~T.,    {Nordlund} A.~P.,  1997, \apj, 474, 730

\bibitem[\protect\citeauthoryear{{Padoan} \& {Nordlund}}{{Padoan} \&
  {Nordlund}}{2002}]{Padoan2002}
{Padoan} P.,  {Nordlund} {\AA}.,  2002, \apj, 576, 870

\bibitem[\protect\citeauthoryear{{Peters}, {Mac Low}, {Banerjee}, {Klessen} \&
  {Dullemond}}{{Peters} et~al.}{2010}]{Peters2010}
{Peters} T.,  {Mac Low} M.-M.,  {Banerjee} R.,  {Klessen} R.~S.,    {Dullemond}
  C.~P.,  2010, \apj, 719, 831

\bibitem[\protect\citeauthoryear{{Price}}{{Price}}{2007}]{Price2007}
{Price} D.~J.,  2007, Publ. Astron. Soc. Aust., 24, 159

\bibitem[\protect\citeauthoryear{{S{\'a}nchez}, {Alfaro} \&
  {P{\'e}rez}}{{S{\'a}nchez} et~al.}{2005}]{Sanchez2005}
{S{\'a}nchez} N.,  {Alfaro} E.~J.,    {P{\'e}rez} E.,  2005, \apj, 625, 849

\bibitem[\protect\citeauthoryear{{Sandford} II, {Whitaker} \&
  {Klein}}{{Sandford} et~al.}{1982}]{Sandford1982}
{Sandford} II M.~T.,  {Whitaker} R.~W.,    {Klein} R.~I.,  1982, \apj, 260, 183

\bibitem[\protect\citeauthoryear{{Shadmehri} \& {Elmegreen}}{{Shadmehri} \&
  {Elmegreen}}{2011}]{Shadmehri2011}
{Shadmehri} M.,  {Elmegreen} B.~G.,  2011, \mnras, 410, 788

\bibitem[\protect\citeauthoryear{{Spitzer}}{{Spitzer}}{1978}]{Spitzer1978}
{Spitzer} L.,  1978, {Physical processes in the interstellar medium}

\bibitem[\protect\citeauthoryear{{Springel}, {Yoshida} \& {White}}{{Springel}
  et~al.}{2001}]{Springel2001}
{Springel} V.,  {Yoshida} N.,    {White} S.~D.~M.,  2001, New Astronomy, 6, 79

\bibitem[\protect\citeauthoryear{{Stamatellos}, {Whitworth} \&
  {Hubber}}{{Stamatellos} et~al.}{2011}]{Stamatellos2011}
{Stamatellos} D.,  {Whitworth} A.~P.,    {Hubber} D.~A.,  2011, \apj, 730, 32

\bibitem[\protect\citeauthoryear{{Stutzki}, {Bensch}, {Heithausen}, {Ossenkopf}
  \& {Zielinsky}}{{Stutzki} et~al.}{1998}]{Stutzki1998}
{Stutzki} J.,  {Bensch} F.,  {Heithausen} A.,  {Ossenkopf} V.,    {Zielinsky}
  M.,  1998, \aap, 336, 697

\bibitem[\protect\citeauthoryear{{Thompson}, {Urquhart}, {Moore} \&
  {Morgan}}{{Thompson} et~al.}{2012}]{Thompson2012}
{Thompson} M.~A.,  {Urquhart} J.~S.,  {Moore} T.~J.~T.,    {Morgan} L.~K.,
  2012, \mnras, 421, 408

\bibitem[\protect\citeauthoryear{{Tremblin}, {Audit}, {Minier}, {Schmidt} \&
  {Schneider}}{{Tremblin} et~al.}{2012}]{Tremblin2012}
{Tremblin} P.,  {Audit} E.,  {Minier} V.,  {Schmidt} W.,    {Schneider} N.,
  2012, \aap, 546, A33

\bibitem[\protect\citeauthoryear{{Vogelaar} \& {Wakker}}{{Vogelaar} \&
  {Wakker}}{1994}]{Vogelaar1994}
{Vogelaar} M.~G.~R.,  {Wakker} B.~P.,  1994, \aap, 291, 557

\bibitem[\protect\citeauthoryear{{Walch}, {Whitworth}, {Bisbas}, {W{\"u}nsch}
  \& {Hubber}}{{Walch} et~al.}{2013}]{Walch2013}
{Walch} S.,  {Whitworth} A.~P.,  {Bisbas} T.~G.,  {W{\"u}nsch} R.,    {Hubber}
  D.~A.,  2013, \mnras, 435, 917

\bibitem[\protect\citeauthoryear{{Walch}, {Whitworth} \& {Girichidis}}{{Walch}
  et~al.}{2011}]{Walch2011}
{Walch} S.,  {Whitworth} A.~P.,    {Girichidis} P.,  2011, \mnras, pp accepted
  by MNRAS, arXiv:1109.0280

\bibitem[\protect\citeauthoryear{{Walch}, {Whitworth}, {Bisbas}, {W{\"u}nsch}
  \& {Hubber}}{{Walch} et~al.}{2012}]{Walch2012}
{Walch} S.~K.,  {Whitworth} A.~P.,  {Bisbas} T.,  {W{\"u}nsch} R.,    {Hubber}
  D.,  2012, \mnras, 427, 625

\bibitem[\protect\citeauthoryear{{White}, {Nelson}, {Holland}, {Robson},
  {Greaves}, {McCaughrean}, {Pilbratt}, {Balser}, {Oka}, {Sakamoto},
  {Hasegawa}, {McCutcheon}, {Matthews}, {Fridlund}, {Tothill}, {Huldtgren} \&
  {Deane}}{{White} et~al.}{1999}]{White1999}
{White} G.~J.,  {Nelson} R.~P.,  {Holland} W.~S.,  {Robson} E.~I.,  {Greaves}
  J.~S.,  {McCaughrean} M.~J.,  {Pilbratt} G.~L.,  {Balser} D.~S.,  {Oka} T.,
  {Sakamoto} S.,  {Hasegawa} T.,  {McCutcheon} W.~H.,  {Matthews} H.~E.,
  {Fridlund} C.~V.~M.,  {Tothill} N.~F.~H.,  {Huldtgren} M.,    {Deane} J.~R.,
  1999, \aap, 342, 233

\bibitem[\protect\citeauthoryear{{Whitworth}, {Bhattal}, {Chapman}, {Disney} \&
  {Turner}}{{Whitworth} et~al.}{1994a}]{Whitworth1994b}
{Whitworth} A.~P.,  {Bhattal} A.~S.,  {Chapman} S.~J.,  {Disney} M.~J.,
  {Turner} J.~A.,  1994a, \aap, 290, 421

\bibitem[\protect\citeauthoryear{{Whitworth}, {Bhattal}, {Chapman}, {Disney} \&
  {Turner}}{{Whitworth} et~al.}{1994b}]{Whitworth1994a}
{Whitworth} A.~P.,  {Bhattal} A.~S.,  {Chapman} S.~J.,  {Disney} M.~J.,
  {Turner} J.~A.,  1994b, \mnras, 268, 291

\bibitem[\protect\citeauthoryear{{W{\"u}nsch}, {Dale}, {Palou{\v s}} \&
  {Whitworth}}{{W{\"u}nsch} et~al.}{2010}]{Wunsch2010}
{W{\"u}nsch} R.,  {Dale} J.~E.,  {Palou{\v s}} J.,    {Whitworth} A.~P.,  2010,
  \mnras, 407, 1963

\bibitem[\protect\citeauthoryear{{Zanstra}}{{Zanstra}}{1951}]{Zanstra1951}
{Zanstra} H.,  1951, \bain, 11, 359

\bibitem[\protect\citeauthoryear{{Zavagno}, {Deharveng}, {Comer{\'o}n},
  {Brand}, {Massi}, {Caplan} \& {Russeil}}{{Zavagno}
  et~al.}{2006}]{Zavagno2006}
{Zavagno} A.,  {Deharveng} L.,  {Comer{\'o}n} F.,  {Brand} J.,  {Massi} F.,
  {Caplan} J.,    {Russeil} D.,  2006, \aap, 446, 171

\bibitem[\protect\citeauthoryear{{Zavagno}, {Pomar{\`e}s}, {Deharveng},
  {Hosokawa}, {Russeil} \& {Caplan}}{{Zavagno} et~al.}{2007}]{Zavagno2007}
{Zavagno} A.,  {Pomar{\`e}s} M.,  {Deharveng} L.,  {Hosokawa} T.,  {Russeil}
  D.,    {Caplan} J.,  2007, \aap, 472, 835

\bibitem[\protect\citeauthoryear{{Zavagno}, {Russeil}, {Motte}, {Anderson},
  {Deharveng}, {Rod{\'o}n}, {Bontemps}, {Abergel}, {Baluteau}, {Sauvage},
  {Andr{\'e}}, {Hill} \& {White}}{{Zavagno} et~al.}{2010}]{Zavagno2010}
{Zavagno} A.,  {Russeil} D.,  {Motte} F.,  {Anderson} L.~D.,  {Deharveng} L.,
  {Rod{\'o}n} J.~A.,  {Bontemps} S.,  {Abergel} A.,  {Baluteau} J.-P.,
  {Sauvage} M.,  {Andr{\'e}} P.,  {Hill} T.,    {White} G.~J.,  2010, \aap,
  518, L81

\end{thebibliography}


\end{document}